# Observation of the orbital Hall effect in a light metal Ti

Young-Gwan Choi[1†], Daegeun Jo[2†], Kyung-Hun Ko[1†], Dongwook Go[3,4], Kyung-Han Kim[2], Hee Gyum Park[5], Changyoung Kim[6,7], Byoung-Chul Min[5], Gyung-Min Choi[1,8★], and Hyun-Woo Lee[2,9★]

[1]Department of Energy Science, Sungkyunkwan University, Suwon 16419, Korea
[2]Department of Physics, Pohang University of Science and Technology, Pohang 37673, Korea
[3]Peter Grünberg Institut and Institute for Advanced Simulation, Forschungszentrum Jülich and JARA, 52425 Jülich, Germany
[4]Institute of Physics, Johannes Gutenberg University Mainz, 55099 Mainz, Germany
[5]Center for Spintronics, Korea Institute of Science and Technology, Seoul 02972, Korea
[6]Department of Physics and Astronomy, Seoul National University, Seoul 08826, Korea
[7]Center for Correlated Electron Systems, Institute for Basic Science, Seoul 08826, Korea
[8]Center for Integrated Nanostructure Physics, Institute for Basic Science, Suwon 16419, Korea
[9]Asia Pacific Center for Theoretical Physics, 77 Cheongam-ro, Pohang 37673, Korea

[†]These authors equally contributed to this work.
★Correspondence to: gmchoi@skku.edu (G.-M. Choi), hwl@postech.ac.kr (H.-W. Lee)

The orbital angular momentum is a core ingredient of orbital magnetism[1], spin Hall effect[2,3], giant Rashba spin splitting[4,5], orbital Edelstein effect[6], and spin-orbit torque[7,8]. However, its experimental detection is tricky. In particular, direct detection of the orbital Hall effect[9] remains elusive[10-13] despite its importance[2,14] for electrical control of magnetic **nanodevices. Here we report the direct observation of the orbital Hall effect in a light metal Ti. The Kerr rotation by the accumulated orbital magnetic moment is measured at Ti surfaces, whose result agrees with theoretical calculations semiquantitatively and is supported by the orbital torque measurement[15] in Ti-based magnetic heterostructures. The results confirm the electron orbital angular momentum as an essential dynamic degree of freedom[9,16-20], which may provide a novel mechanism for the electric control of magnetism[21,22]. The results may also deepen the understanding of spin[2,3,14], valley[23,24], phonon[25-27], and magnon[28,29] dynamics coupled with orbital dynamics.**

Fast electrical control of magnetism can be achieved by injecting a spin current into nanomagnets[8,30-32]. The spin Hall effect (SHE)[33-35] generates a spin current perpendicular to an external electric field by inducing spin-dependent transverse motion of electrons (Fig. 1a). Search for materials with stronger SHE is in progress to realize SHE-based nonvolatile memory devices[32,35]. Theoretical studies[2,3,14] indicate that the SHE arises in $4d$ and $5d$ transition metals[2,14] and possibly many other materials[3,9,18,36] due to the orbital Hall effect (OHE)[3,9], which induces orbital-angular-momentum-dependent transverse flows (Fig. 1b). The OHE-origin of the SHE explains[2,14] the systematic variation of the spin Hall conductivity sign in $5d$ transition metals[31,35,37]. However, direct experimental evidence of the OHE is lacking[10-13].

The most direct observation of the OHE is arguably to measure orbital moments accumulated at surfaces of OHE-exhibiting materials. There are a few problems to overcome. Firstly, the OHE and the SHE usually go together, and distinguishing the resulting orbital and spin signals from each other is challenging since their properties are similar[2,3]. To overcome this problem, we utilize the fact that the spin-orbit coupling is compulsory for the SHE but not for the OHE[3,36]. Thus, in light metals with negligible spin-orbit coupling, the SHE becomes negligible but the OHE can remain strong provided the orbital character of wave functions varies fast with crystal momentum (orbital texture[3,36]). We thus measure the orbital moment accumulation for the light metal Ti using the magneto-optical Kerr effect (MOKE). Our theoretical calculations (SI Sec. 1) verify that fcc Ti has pronounced orbital texture (Fig. 1c) and large orbital Berry curvature (Fig. 1d), which results in large orbital Hall conductivity $\sigma_{\text{OH}}$ [3800 $(\hbar/e)(\Omega \cdot \text{cm})^{-1}$] that is two orders of magnitude larger than its spin Hall conductivity $\sigma_{\text{SH}}$ [-40 $(\hbar/e)(\Omega \cdot \text{cm})^{-1}$] (Fig. 1e). Another problem is the orbital quenching by the crystal field, which may severely suppress the orbital accumulation. Fortunately, the MOKE is two orders of magnitude more sensitive to the orbital magnetization than the spin magnetization in Ti (lower panel in Fig. 1e; see also SI Sec. 2) since the spin-orbit coupling, crucial for the sensitivity to the spin magnetization, is weak in Ti[38]. Thus, the MOKE signal of Ti is still dominated by the orbital accumulation even when the crystal field suppresses it. As an independent test, we also measure the orbital torque[15] that arises when an OHE-induced orbital current is injected into an adjacent ferromagnet.

Figure 2a shows the MOKE measurement geometry. When a charge current flows along the *x*-direction, the OHE results in the orbital accumulation at the top surface of Ti, whose magnetization is along the *y*-direction. To measure the *y*-magnetized orbitals, we use the longitudinal MOKE (LMOKE) geometry: a laser beam with *s*-polarization is incident at the center of the top surface with the incidence angle of $\phi = 25°$ in the *yz*-plane (Methods and SI Sec. 3). The measured Kerr rotation angle $\theta_K$ (Fig. 2b) for the beam reflected at the center of the Ti top surface (Ti thickness 90 nm) grows linearly with the charge current density $j_c$ and changes its sign for the opposite incidence angle of $\phi = -25°$, which confirm that $\theta_K$ is not due to the Joule heating[39] but instead originates from the *y*-polarized orbitals. For comparison, $\theta_K$ is measured for a Pt control sample (Pt thickness 70 nm, SI Sec. 4). Interestingly, $\theta_K$ per $j_c$ is comparable for Ti and Pt; $\theta_K/j_c \approx -48.3\pm20.4$ nrad/($10^7$ A/cm$^2$) for Ti and $+24.3\pm4.0$ nrad/($10^7$ A/cm$^2$) for Pt (Fig. 2c). The latter value is comparable to the value $+50$ nrad/($10^7$ A/cm$^2$) obtained previously[40] with a different laser frequency (Methods) and attributed to the SHE. In contrast, large $\theta_K$ for Ti cannot be attributed to the SHE since the SHE is quite weak and the MOKE is insensitive to the spin accumulation in Ti.

The Oersted field could generate $\theta_K$ linearly proportional to $j_c$. To investigate this possibility, $\theta_K$ is scanned along the $y$-direction for $\phi = \pm 25°$ and decomposed into the LMOKE signal $\theta_K^L = [\theta_K(\phi=+25°) - \theta_K(\phi=-25°)]/2$, which probes the $y$-component of magnetization, and the polar MOKE (PMOKE) signal $\theta_K^P = [\theta_K(\phi=+25°) + \theta_K(\phi=-25°)]/2$, which probes the $z$-component of magnetization (SI Sec. 5). $\theta_K^L$ for Ti is essentially independent of $y$ (black circles in Fig. 2d), which is the expected behavior for the current-induced orbital accumulation and similar to the SHE-driven $\theta_K^L$ for Pt (Fig. 2e). On the other hand, $\theta_K^P$ for Ti is antisymmetric in $y$ and follows the spatial profile (solid red line in Fig. 2d) expected for the accumulation induced by the Oersted field $z$ component ($\theta_K^P$ for Pt is negligible, SI Sec. 6). To check if $\theta_K^P$ for Ti is due to the Oersted field, we calculate the Oersted-field-induced orbital magnetization (SI Sec. 6) and the resulting $\theta_K^P$. It turns out that the computed peak values $\pm 23$ nrad/($3\times10^6$ A/cm$^2$) at the edges ($y=\pm10$ $\mu$m) are in reasonable agreement with the measured values $\theta_K^P = \pm 19$ nrad/($3\times10^6$ A/cm$^2$). On the other hand, a similar calculation (SI Sec. 6) of $\theta_K^L$ originating from the Oersted-field-induced orbital magnetization turns out to be one order of magnitude smaller than the measured value. We thus conclude that $\theta_K^L$ is not due to the Oersted field whereas $\theta_K^P$ arises from the Oersted field in the Ti film (thickness 90 nm).

To investigate the relationship between $\theta_K^L$ and the OHE, we measure both $\theta_K^L$ and the Kerr ellipticity $\varepsilon_K^l$ as a function of the Ti film thickness $t$=9~90 nm (Fig. 2f). Interestingly, $\theta_K^L$ changes its sign around $t$=45 nm, which is in clear contrast to the monotonic increase of $\theta_K^L$ with $t$ for Pt [40]. On the other hand, $\varepsilon_K^l$ for Ti increases monotonically with $t$. To understand the $t$-dependences (SI Secs.7-9), the complex Kerr rotation $\tilde{\theta}_K^L = \theta_K^L + i\varepsilon_K^l$ is expressed as

$$\tilde{\theta}_K^L \approx \left(\tilde{\theta}_K^L\right)_\infty^{\text{unit}} \int_0^t M_y(z) \kappa e^{-\kappa z} dz, \tag{1}$$

where $\left(\tilde{\theta}_K^L\right)_\infty^{\text{unit}}$ denotes the complex Kerr rotation per uniform orbital magnetization in bulk Ti ($t$=∞) in Fig. 1e, $M_y(z)$ is the y-component of the magnetization at depth $z$ of the film ($z$=0 for the top surface and $z$=$t$ for the bottom surface), and $\kappa$= (0.58−$i$0.56)×10$^8$ m$^{-1}$ is determined from the independently measured refractive index $n$ (SI Sec. 11). For quantitative analysis, we use the orbital drift-diffusion equation and obtain the depth profile of magnetization (SI Sec. 7),

$$M_y(z) = M_0 \frac{\sinh[(z - t/2)/l_L]}{\cosh(t/2l_L)}, \tag{2}$$

with $M_0 = \gamma_L \sigma_{\text{OH}}^{\text{pr}} j_c \rho l_L / D_L$. Here the electrical resistivity $\rho$ is measured independently (SI Sec. 11). Equation (2) and the expression for $M_0$ are similar to the SHE-induced spin accumulation profile in Pt[40] except for the trivial replacement of spin variables by the corresponding orbital variables such as the orbital relaxation length $l_L$, the gyromagnetic ratio $\gamma_L$ for orbital, and the orbital diffusion constant $D_L$. There is a nontrivial replacement, however: the *proper* orbital Hall conductivity $\sigma_{\text{OH}}^{\text{pr}}$ appears in $M_0$ instead of the conventional orbital Hall conductivity $\sigma_{\text{OH}}$. $\sigma_{\text{OH}}^{\text{pr}}$ is the orbital version of the proper spin Hall conductivity[41] and differs from $\sigma_{\text{OH}}$ in that $\sigma_{\text{OH}}^{\text{pr}}$ takes account of the intrinsic orbital relaxation by the crystal field (SI Sec. 7).

Equations (1) and (2) are used to fit the measured $t$-dependences of $\theta_K^L$ and $\varepsilon_K^L$. Since the fitting is sensitive to the phase angle of $(\tilde{\theta}_K^L)_\infty^{\text{unit}}$, we use $\arg\left[(\tilde{\theta}_K^L)_\infty^{\text{unit}}\right]$, $l_L$, and $M_0$ as free parameters, whereas the independently evaluated $\left|(\tilde{\theta}_K^L)_\infty^{\text{unit}}\right|$ (SI Sec. 2) and $\kappa$ remain fixed. Both $\theta_K^L$ and $\varepsilon_K^L$ can be fitted excellently (solid lines in Fig.2f) for $l_L = 74 \pm 24$ nm, $\arg\left[(\tilde{\theta}_K^L)_\infty^{\text{unit}}\right] = -0.95 \pm 0.02$ rad, and $M_0 = (3.5 \pm 0.7) \times 10^{-5}$ $\mu_B$/atom per $j_c$ of $10^7$ A/cm$^2$. The fitted value of $l_L$ agrees with the orbital relaxation length ($61 \pm 18$ nm) obtained from the orbital torque measurement (Fig. 3f) presented below. The fitted value of $\arg\left[(\tilde{\theta}_K^L)_\infty^{\text{unit}}\right]$ agrees reasonably well with the theoretical value $\arg[(-3.8 + i7.3) \times 10^{-3}$ rad/$(\mu_B$/atom$)] = -1.1$ rad for the orbital magnetization (lower panel in Fig. 1e). We then use the relation $M_0 = \gamma_L \sigma_{\text{OH}}^{\text{pr}} j_c \rho l_L / D_L$ to convert the fitted value of $M_0$ to $\sigma_{\text{OH}}^{\text{pr}}$. $D_L$ is expected to be comparable to the charge diffusion constant $D$, but its precise value is unknown. Assuming that $D_L$ equals $D = \left(e^2 \rho N(E_F)\right)^{-1}$ with the measured value of $\rho$ and the calculated value of the density of states $N(E_F)$, we obtain $\sigma_{\text{OH}}^{\text{pr}} = 36 \pm 7$ $(\hbar/e)(\Omega \cdot \text{cm})^{-1}$, which is about five times smaller than the theoretically estimated value $\sigma_{\text{OH}}^{\text{pr}} = 200$ $(\hbar/e)(\Omega \cdot \text{cm})^{-1}$ (SI Sec. 7). Considering the uncertainty in the ratio $D_L/D$, we argue that the fitted and the theoretically estimated values are in reasonable agreement. Note that $\sigma_{\text{OH}}^{\text{pr}}$ is significantly smaller than the conventional orbital Hall conductivity $\sigma_{\text{OH}} = 3800$ $(\hbar/e)(\Omega \cdot \text{cm})^{-1}$, implying that the orbital quenching by the crystal field severely suppresses the orbital accumulation. By the way, if we assume that $\theta_K^L$ and $\varepsilon_K^L$ arise from the spin accumulation caused by the SHE in Ti, we obtain $M_0 = -2.8 \times 10^{-3}$ $\mu_B$/atom per $10^7$ A/cm$^2$ and $\sigma_{\text{SH}} = -1400$ $(\hbar/e)(\Omega \cdot \text{cm})^{-1}$, both of which are unrealistically large, implying that the SHE is not responsible for the measured Kerr signals.

We thus conclude that the OHE is the more probable origin of the Kerr signals than the SHE (see SI Secs. 8 & 9 for other fitting methods).

Next, we present experimental evidence of the orbital torque[15] arising from an OHE-induced orbital current injected into a ferromagnet. We measure current-induced torque in Ti ($t$)/CoFeB (3 nm) bilayers (Fig. 3a) using the quadratic MOKE (QMOKE) in a polar geometry ($\phi$=0) (SI Sec. 12). When the magnetization of CoFeB is tilted by the torque away from the equilibrium $x$-direction set by an external magnetic field, the resulting Kerr rotation by CoFeB is expressed[42] as $\Delta\theta_K = \alpha_{MO}\Delta M_z + \beta_{MO}\cos(2\psi)\Delta M_y$, where $\alpha_{MO}$ and $\beta_{MO}$ are, respectively, linear and quadratic MOKE coefficients of CoFeB, and $\psi$ is the angle between light's polarization and the magnetic field (Fig. 3a). $\alpha_{MO}$ and $\beta_{MO}$ of each sample are determined from separate measurements (SI Secs. 13,14). The orbital torque from Ti induces the damping-like ($h_{DL}$) and field-like ($h_{FL}$) effective fields, which are responsible for $\Delta M_z$ and $\Delta M_y$, respectively, and can be distinguished by their dependence on $\psi$ (Fig. 3b). In addition to the orbital torque, the Oersted field also induces a torque on CoFeB. We predetermine the Oersted field from the spatial variation of $\Delta\theta_K$ along the channel width (Fig. 3c). Then, from the magnetic field dependence of $\Delta M_y$ and $\Delta M_z$, we find that the orbital torque induces a sizable $h_{FL}$ but negligible $h_{DL}$ (Figs. 3d & 3e). Defining a normalized torque efficiency $\xi^j_{DL/FL} = \frac{2e}{\hbar}M_s t_F \frac{h_{DL/FL}}{j_c}$, where $M_s$ is the saturation magnetization of CoFeB, $t_F$ is the thickness of CoFeB, and $j_c$ is the current density in Ti, we obtain $\xi^j_{FL} = -0.32\pm0.06$ and $\xi^j_{DL} = 0.01\pm0.01$ in the Ti (100 nm)/CoFeB (3 nm) sample.

To investigate the origin of $\xi^j_{FL/DL}$, the $t$ (Ti thickness) dependences of $\xi^j_{FL/DL}$ are measured (Fig. 3f). While $\xi^j_{DL}$ stays negligible regardless of $t$, $\xi^j_{FL}$ increases with $t$, which implies that the field-like torque arises from the Ti *bulk* rather than the Ti/CoFeB interface. The measured field-like torque cannot be attributed to the bulk SHE in Ti since its $\sigma_{SH}$ is tiny. Moreover, the spin current injection usually produces the damping-like torque dominantly instead of the field-like torque[43]. For the OHE, on the other hand, the resulting torque can be primarily damping-like or field-like depending on material details, according to the theoretical calculation (Fig. 4a vs. Fig. S2 in ref.[44]). The orbital torque calculation for a Ti/Co bilayer (SI Sec. 15) finds primarily field-like torque of negative sign, in agreement with the experimental result in Fig. 3f. We thus attribute the measured field-like torque to the bulk OHE in Ti. To examine this interpretation further, we fit the $t$ dependence of $\xi_{FL}^j$ by the expression,

$$\xi_{FL}^j = \theta_{OH}^{FL}\left(1 - \text{sech}\left(\frac{t}{l_L}\right)\right) + \xi_{int}^j, \tag{3}$$

where $\theta_{OH}^{FL}$ is the effective orbital Hall angle for the field-like torque, and $\xi_{int}^j$ is the torque efficiency of interfacial origin. From the fitting (Fig. 3f), we obtain $l_L = 61\pm18$ nm, $\theta_{OH}^{FL} = -0.36 \pm 0.07$, and $\xi_{int}^j = 0.04\pm0.03$. $l_L$ matches excellently the value (74±24 nm) obtained from the orbital accumulation measurement (Fig. 2f), providing quantitative evidence that the measured field-like torque arises from the OHE in Ti. The value of $\theta_{OH}^{FL}$ is larger than the spin Hall angles of heavy metals such as Pt, Ta, and comparable to that of W [refs. 31,35,37]. From $\theta_{OH}^{FL}$, we evaluate the effective orbital Hall conductivity for the field-like torque, $\sigma_{OH}^{FL} = \theta_{OH}^{FL}/2\rho = 2000\ (\hbar/e)(\Omega\cdot\text{cm})^{-1}$, which is in order of magnitude agreement with the theoretically calculated orbital Hall conductivity $\sigma_{OH}$ [3800 $(\hbar/e)(\Omega\cdot\text{cm})^{-1}$] but one order of magnitude larger than the proper orbital Hall conductivity $\sigma_{OH}^{pr}$ [200 $(\hbar/e)(\Omega\cdot\text{cm})^{-1}$] that takes into account the orbital relaxation by the crystal field. This result implies that the crystal field severely suppresses only the orbital accumulation but not the orbital current. A similar result is obtained from the theoretical calculation (SI Sec. 15).

Recent experiments[13,21] suggested that an orbital-current-induced torque becomes more damping-like when a thin Pt layer is inserted in bilayer systems. We find a similar result for a Ti (100 nm)/Pt (3 nm)/CoFeB (3 nm) trilayer: The Pt layer insertion greatly enhances the damping-like torque efficiency $\xi_{DL}^{j}$ from 0.01 to 0.13±0.03 (Fig. 3f). The enhanced value is significantly larger than the corresponding efficiency 0.065 for Pt(10 nm)/CoFeB(3 nm) (SI Sec. 14) and thus cannot be attributed to the SHE in Pt. Our torque calculation for a Ti/Pt/Co trilayer indicates that the enhanced damping-like torque originates from the OHE in Ti (SI Sec. 15).

Our experimental results confirm the OHE in a light metal Ti and hint at opportunities in the emerging field of orbitronics[9,16,17]. Recently reported large torques from the orbital current[21,22,45] without any heavy metals widen material choice for the electrical control of magnetism. Sizable orbital relaxation length in Ti implies that the orbital may be used as an information carrier and explains why recent experiments[46,47] on very thin Ti failed to observe the OHE. Our results also reveal interesting differences between the orbital and spin dynamics, which require an improved understanding of orbital dynamics[48]. The orbitronics also deepens understanding of other degrees of freedom since the orbital degree of freedom may play pivotal roles for the spin dynamics through the spin-orbit coupling[2,3,14], the valley dynamics through the orbital dependence of valleys[23,24], the phonon dynamics[25] through the orbital-lattice deformation coupling[26,49], and the magnon dynamics through the spin-orbit coupling[28] or the scalar spin chirality[29]. More exotic opportunities may arise by linking the orbital and other dynamics.

**Methods**

*Device fabrication and characterization*: Ti, Pt, and Ti/CoFeB samples are deposited onto sapphire substrates using the magnetron dc sputtering technique. The sputtering conditions are the base pressure of $10^{-8}$ Torr, process pressure of 5 mTorr of Ar, and substrate temperature of 300 K. Deposition rates of Ti, Pt, and CoFeB are 0.039, 0.218, and 0.066 nm/s, respectively. To prevent the oxidation of deposited films, we deposit a 3-nm SiN layer on top of the metallic layers without breaking vacuum conditions. Devices are patterned by the photolithography method with bow-tie shape. The channel width at the center region is 20 $\mu$m for Ti and 10 $\mu$m for Pt. We check the fcc crystal structure of Ti by using X-ray diffraction (SI Sec. 10). We determine the refractive indices of Ti, Pt, and CoFeB in the wavelength range of 200~1000 nm using ellipsometry (SI Sec. 11). We measure the electrical resistivity of $9.0\times10^{-7}$, $3.0\times10^{-7}$, and $12\times10^{-7}$ Ohm·m of Ti, Pt, and CoFeB, respectively, by using the four-point probe method (SI Sec. 11).

*Detection of orbital accumulation using longitudinal-linear MOKE*: By injecting an AC current on the Ti single layer, we induce the orbital accumulation on the top surface of Ti. To detect the orbital accumulation, a linearly polarized light with a wavelength of 780 nm is incident on the Ti surface with an oblique incident angle of $\pm 25°$ (longitudinal geometry). The orbital accumulation results in the Kerr rotation of the reflected light, which is linearly proportional to the current density and changes its sign with the opposite incident angle. The magnitude of the Kerr rotation is determined by the linear magneto-optic constant and refractive index of Ti, the thickness profile of the orbital accumulation, and the incident angle of light (SI Sec. 5). Sensitive measurement of the Kerr rotation with a noisy level of <10 nrad is achieved by combining AC current source, lock-in amplifier, and balanced photodetector (SI Secs. 3 and 5).

*Detection of orbital torque using polar-quadratic MOKE*: By injecting an AC current on the Ti/CoFeB, Ti/Pt/CoFeB, or Pt/CoFeB heterostructures, we induce the orbital torque on the CoFeB magnetization. To detect the orbital torque, a linearly polarized light with a wavelength of 780 nm is incident on the CoFeB surface with an incident angle 0º (polar geometry) (SI Sec. 12). The field-like torque, driven by the orbital current from Ti to CoFeB, tilts the CoFeB magnetization, resulting in the Kerr rotation of the reflected light. The magnitude of the Kerr rotation is determined by the quadratic magneto-optic constant of CoFeB, refractive indexes of Ti and CoFeB, initial polarization of light, and applied magnetic field (SI Secs. 13,14). With the same geometry, the Kerr rotation from the damping-like torque is determined by the linear magneto-optic constant of CoFeB and independent of the initial polarization of light and applied magnetic field (SI Sec. 14).

**Acknowledgments**

D.J. was supported by the Global Ph.D. Fellowship Program by National Research Foundation of Korea (Grant No. 2018H1A2A1060270). D.J., K.-H.K., and H.-W.L. were supported by the Samsung Science and Technology Foundation (BA-1501-51). G.-M.C was supported by the National Foundation of Korea (2019R1C1C1009199). H.G.P and B.-C.M were supported by the KIST institutional program and the National Research Foundation of Korea (NRF) programs (2019M3F3A1A02071509 and 2020M3F3A2A01081635).

**Author contributions**

Y.-G.C., D.J., and K.-H.K equally contributed to this work. C.K. initiated the project, and H.-W.L and G.-M.C. supervised the study. Y.-G.C. performed the measurement and analysis for the orbital accumulation in Ti and Pt films. K.-H.K performed the measurement and analysis for the orbital torque in Ti/CoFeB, Ti/Pt/CoFeB, and Pt/CoFeB heterostructures. D.J. and D.G. prepared theoretical formulation of the MOKE analysis. D.J. carried out the tight-binding calculations of the spin and orbital Hall conductivities of bulk Ti, analyzed the MOKE signals theoretically, and calculated the current-induced torque in Ti/Co bilayers. K.-H.K. calculated the proper orbital Hall conductivity of Ti. K.-H.K, Y.-G.C., H.G.P., and B.-C.M fabricated samples and characterized the optical, electrical, and magnetic properties.

**Competing interests**

The authors declare that they have no competing interests.

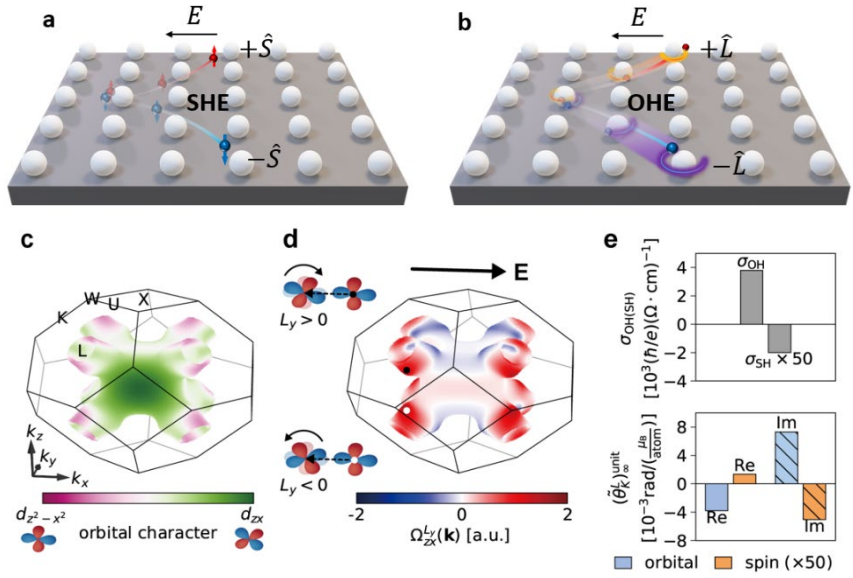

**Figure 1. Schematic illustrations of spin Hall effect (SHE), orbital Hall effect (OHE), and its simulations for Ti. a,b,** When electrons flow due to an electric bias field, there occurs spin-dependent deflection motion making a spin current along transverse direction inducing the accumulation of spin angular momentum $\hat{S}$ at boundaries (SHE). For the OHE, electrons with orbital angular momentum $\hat{L}$ can exhibit deflection motion even without spin-orbit coupling, thereby inducing the accumulation of orbital angular momentum at boundaries. **c,** The orbital texture on the inner Fermi surface of fcc Ti from the realistic tight-binding calculations (SI Sec. 1 for calculation details and results on the outer Fermi surface). The color implies whether the wave function is more like $d_{z^2-x^2}$ than $d_{zx}$ (pink) or vice versa (green) for each crystal momentum $k$. **d,** $k$-resolved orbital Hall conductivity under an electric field along the $x$-direction. The orbital Hall conductivity is large where the orbital texture is strong. The black and white points denote two particular $k$ points with the same $d_{z^2-x^2}$ character but with opposite signs of $k_z$. They evolve to the finite orbital angular momentum states $|L_y = \pm 2\hbar\rangle \propto |d_{z^2-x^2}\rangle \pm i|d_{zx}\rangle$, which give rise to the positive orbital current $\sim k_z L_y$. **e,** The orbital and spin Hall conductivities (upper panel) of fcc Ti and the longitudinal complex Kerr angle (lower panel) from the uniform orbital or spin magnetization (real and imaginary parts of the complex Kerr angle plotted separately). The quantities for orbital are about two orders of magnitude larger than those for spin.

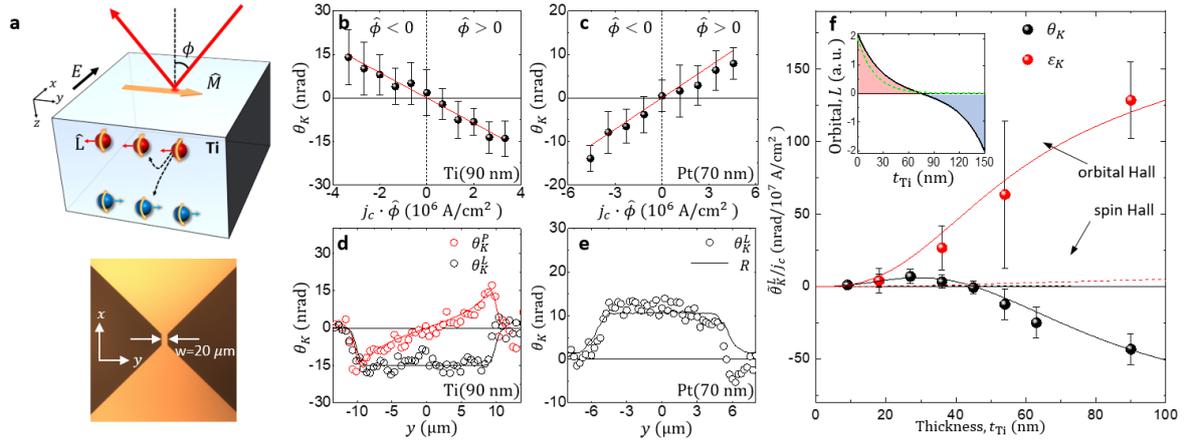

**Figure 2. Measurement of orbital accumulation**. **a**, Schematic illustration of longitudinal MOKE experiment (upper panel) and an optical image of a bow-tie shaped device (lower panel) for Ti film. $E$ is an electric field, $\hat{L}$ is the orbital angular momentum, $\hat{M}$ is the orbital magnetization, $\phi$ is a light incidence angle, and $\hat{\theta}_K$ is the complex Kerr rotation. **b,c**, MOKE signal for Ti with thickness 90 nm (b) and Pt with thickness 70 nm (c) as a function of current density, $j_c$. The magnitude of MOKE signals is linearly proportional to $j_c$. To remove the Joule heating effect, all data points are collected by subtracting signals measured with opposite polarities of $j_c$. The left side of the origin corresponds to the negative incidence angle of the probing light. **d**, Spatial scanning data of MOKE signal of the Ti 90 nm along the $y$-axis with $j_c$ of $3\times10^6$ A cm$^{-2}$. Black and red data are longitudinal and polar components of Kerr angle, $\theta_K^L$ and $\theta_K^P$, respectively. **e**, Spatial scanning data of MOKE signal of the Pt with thickness 70 nm along the $y$-axis with $j_c$ of $4.6\times10^6$ A cm$^{-2}$. $R$ is the reflected intensity of light. **f**, Thickness dependence of the MOKE signal of the Ti thin films. Black and red circles are the real and imaginary parts of the Kerr rotation. Error bars denote the confidence level of repeated measurements. Solid lines are fitting curves based on the theoretically calculated orbital contributions for MOKE signal with $l_L$ of 74 nm. Dashed lines are fitting curves of the spin contributions calculated by the predetermined parameters (SI Sec. 8). The inset shows an example of a depth profile of the orbital angular momentum accumulation (black solid line) and attenuating light profile (green dashed line). This propagating light into the Ti thin film can contribute to the complicated thickness-dependent MOKE signals.

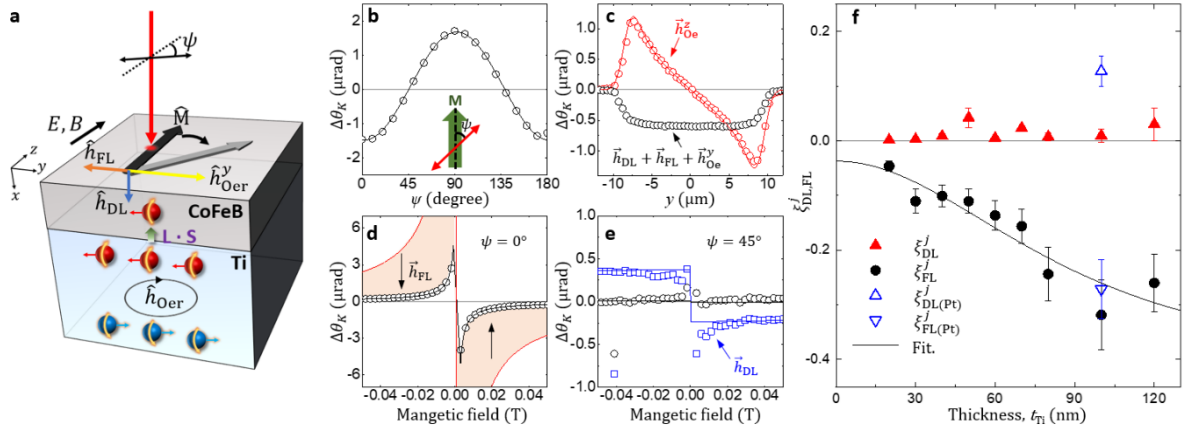

**Figure 3. Measurement of orbital torque a**, Schematic illustration of the magnetization tilt of CoFeB due to the orbital current injected from Ti in the Ti/CoFeB heterostructure. $\psi$ is light polarization angle, $E$ is an electric field, $B$ is an external magnetic field, $\hat{M}$ is the magnetization of CoFeB, $\vec{h}_{Oe}$ is the Oersted field, and $\vec{h}_{DL/FL}$ is the effective field for the damping-like/field-like component of the orbital torque. **b**, $\psi$ dependence of $\Delta\theta_K$ for Ti (100 nm)/CoFeB (3 nm). Quadratic MOKE signal, driven by $\vec{h}_{FL} + \vec{h}^y_{Oe}$, has a $\cos(2\psi)$ dependence, and linear MOKE signal, driven by $\vec{h}_{DL} + \vec{h}^z_{Oe}$, is independent of $\psi$. **c**, Spatial profile of $\Delta\theta_K$ of Ti (100 nm)/CoFeB (3 nm) along the y-direction at $\psi = 0°$. Red squares are magnetization tilt due to the z-component Oersted field ($\vec{h}^z_{Oe}$) obtained by symmetric components under magnetization reversal, $(\Delta\theta_K(+M) + \Delta\theta_K(-M))/2$. The solid red line is a fit of $\vec{h}^z_{Oe}$. Black circles are magnetization tilt due to the orbital torque and y-component Oersted field ($\vec{h}_{DL} + \vec{h}_{FL} + \vec{h}^y_{Oe}$) obtained by asymmetric components under magnetization reversal, $(\Delta\theta_K(+M) - \Delta\theta_K(-M))/2$. The magnetic field is fixed to 0.006 T for **b** and 0.053 T for **c**. **d**, Magnetic field dependence of $\Delta\theta_K$ (experimental, black circles; fit, black solid line) of the Ti (100 nm)/CoFeB (3 nm) at $\psi = 0°$. The black line is a fit of $\vec{h}_{FL} + \vec{h}^y_{Oe}$, and the red line is a fit of $\vec{h}^y_{Oe}$. **e**, Magnetic field dependence of $\Delta\theta_K$ of the Ti (100 nm)/CoFeB (3 nm)/SiN (5 nm) (black circles) and Ti (100 nm)/Pt (3 nm)/CoFeB (3 nm)/SiN (5 nm) (red squares) at $\psi = 45°$. Lines are fits of $\vec{h}_{DL}$. All data of **b**, **c**, **d**, and **e** are obtained with a charge current of 30 mA. **f**, Ti thickness dependence of the torque efficiency of the Ti ($t$ nm)/CoFeB (3 nm) (filled symbols) and Ti (100 nm)/Pt (3 nm)/CoFeB (3 nm) (empty symbols) structures. Circle and triangle points correspond to field-like and damping-like torque efficiency ($\xi_{DL/FL}$), respectively. The solid line is the fit of Eq. (3) with the orbital diffusion length $l_L$=61 nm.